\documentclass[fp,twocolumn]{jpsj3}

\title{Enhanced Spin-Orbit Coupling in a Correlated Metal}

\author{Katsunori Kubo}
\inst{Advanced Science Research Center,
  Japan Atomic Energy Agency, Tokai, Ibaraki 319-1195, Japan}

\date{\today}

\abst{
We investigate the correlation effects on spin-orbit coupling (SOC)
in a two-orbital Hubbard model on a square lattice
by applying the variational Monte Carlo method.
We consider an effective SOC constant $\lambda_{\text{eff}}$
in the one-body part of the variational wave function
and mainly discuss the cases of the electron number per site $n=1$,
that is, quarter filling.
We find that $\lambda_{\text{eff}}$ is proportional to the bare value $\lambda$
and depends on the electron-electron interactions through $(U'-J')$
in a relatively wide parameter range in the paramagnetic (PM) phase,
where $U'$ is the interorbital Coulomb interaction
and $J'$ is the pair hopping interaction.
Increasing the electron-electron interactions in the PM phase
leads to a transition to an effective one-band state,
in which the upper band becomes empty due to the enhanced $\lambda_{\text{eff}}$.
We also construct phase diagrams considering magnetic order.
The carrier doping effect on $\lambda_{\text{eff}}$ is also investigated.
We find that $\lambda_{\text{eff}}$ enhances
in a strongly correlated region around the Mott transition point
and it is necessary to include the correlation effects
beyond the Hartree-Fock approximation to describe the enhanced SOC properly.
}

\begin{document}
\maketitle

\section{Introduction}
Spin-orbit coupling (SOC) is an indispensable component for
describing the magnetism of rare-earth and actinide materials
and has been studied for a long time.
In recent years, the SOC has attracted renewed interest
in the field of condensed matter physics because of the discoveries of 
intriguing phenomena originating from the SOC such as
the spin Hall effect~\cite{Dyakonov1971JETPL,Dyakonov1971PLA,Hirsch1999,
  Murakami2003,Sinova2004,Kato2004,Wunderlich2005}
and the inverse spin Hall effect~\cite{Hirsch1999,
  Saitoh2006,Valenzuela2006,Kimura2007}
in spintronics~\cite{Zutic2004}
and phenomena in topological insulators~\cite{Kane2005PRL1,Bernevig2006PRL,
  Kane2005PRL2,Bernevig2006Science,Konig2007,Roth2009}.
To observe these phenomena, materials containing heavy elements
have been investigated considering the large SOC.

If we can strengthen the effects of the SOC in materials without heavy elements,
the number of candidate materials for the SOC-originating phenomena
will increase.
For this purpose,
the effect of the Coulomb interaction between electrons may be used.
Indeed, the effective SOC has been reported to become
about twice as large as the bare value by the effect of the Coulomb interaction
in Sr$_2$RhO$_4$~\cite{Liu2008}
and Sr$_2$RuO$_4$~\cite{Veenstra2014,Zhang2016,Kim2018,Tamai2019,
  Linden2020,Cao2021}
mainly based on the comparison between
the angle-resolved photoemission spectroscopy experiments
and the band calculations using the density functional theory.

When the effective SOC becomes large in a multiorbital system,
exotic superconductivity may occur.
In a system with orbital degrees of freedom,
even-parity spin-triplet and odd-parity spin-singlet
superconductivities are possible
with antisymmetric orbital states
in principle~\cite{Klejnberg1999,Han2004,Sakai2004,
  Kubo2007,Kubo2008JPSJ,Kubo2008JOAM},
but they would be difficult to realize in actual materials.
In the presence of the SOC,
this even (odd)-parity spin-triplet (-singlet) component
can be mixed with the ordinary even (odd)-parity spin-singlet (-triplet) state
due to the lack of rotational symmetry in the spin
space~\cite{Cvetkovic2013,Veenstra2014,Vafek2017,Yu2018,Cheung2019,Ramires2019,
  Kaba2019,Huang2019,Suh2020,Clepkens2021}.
Thus, under an enhanced effective SOC,
this mixing effect becomes strong
and the exotic superconducting component would play a significant role
in the superconducting properties.

Another intriguing system with a large SOC is Sr$_2$IrO$_4$.
Without the SOC, this 5$d$ electron system is expected to be metallic
since Sr$_2$RhO$_4$,
which is the 4$d$ counterpart and should have a smaller bandwidth,
is a metal.
In the actual Sr$_2$IrO$_4$, the band is split into narrower ones by the SOC
and the Coulomb interaction is sufficiently large
to induce a Mott transition for the narrow effective
band~\cite{Kim2008,Moon2008,Kim2009,Watanabe2010,Wang2011,Qi2012,Watanabe2014}.
This SOC-driven Mott transition may occur in other systems
if the effective SOC is enhanced.

In this study, we investigate the conditions
for the large enhancement of an effective SOC.
For this purpose, we consider a two-orbital model
as a minimal model to include the SOC.
The enhancement of the effective SOC in two- and three-orbital models
has been studied using the dynamical mean-field theory~\cite{Zhang2016,Kim2018,
  Triebl2018,Linden2020,Cao2021,Richter2021},
Gutzwiller approximation~\cite{Bunemann2016},
and Hartree-Fock (HF) approximation~\cite{Liu2021}.
However, systematic investigation of the effective SOC,
such as dependence on the Coulomb interaction, Hund's coupling,
and doping, has not been carried out.

To investigate the enhancement of the effective SOC systematically,
we apply the variational Monte Carlo (VMC) method~\cite{Yokoyama1987}
to the two-orbital model.
Using the VMC method, we can include
the correlation effects beyond the HF approximation.
As a variational wave function,
we consider a wave function with doublon-holon binding factors
[doublon-holon binding wave function (DHWF)]~\cite{Kaplan1982,Yokoyama1990}.
Here, a doublon means a doubly occupied site and a holon means an empty site.
Intersite factors, such as the doublon-holon binding factors,
are essential to discuss correlation effects,
especially
to describe the Mott insulating state~\cite{Yokoyama2002,Capello2006}.
Indeed, it has been shown that the DHWF can describe the Mott transition
for the single-orbital~\cite{Yokoyama2002,Watanabe2006,Yokoyama2006,Onari2007}
and two-orbital~\cite{Koga2006,Takenaka2012,Kubo2021} Hubbard models.
We will show that to properly describe the enhancement of the effective SOC,
it is necessary to include the correlation effects beyond the HF approximation.

\section{Model}\label{model}
We consider a two-orbital Hubbard model given by
$H=H_{\text{kin}}+H_{\text{SOC}}+H_{I}$,
where $H_{\text{kin}}$ is the kinetic energy term,
$H_{\text{SOC}}$ is the SOC term,
and $H_{I}$ is the onsite interaction term.
As the orbital states,
we consider $(p_x,p_y)$ or $(d_{zx},d_{yz})$ orbitals
and denote them as $(x,y)$.
They are written as
\begin{align}
  |x\rangle &= \frac{1}{\sqrt{2}} (-|l 1\rangle +|l -1\rangle), \label{eq:x}\\
  |y\rangle &= \frac{1}{\sqrt{2}i}(-|l 1\rangle -|l -1\rangle), \label{eq:y}
\end{align}
where $l$ is the magnitude of the orbital angular momentum
[$l=1$ for the $(p_x,p_y)$ orbitals and $l=2$ for the $(d_{zx},d_{yz})$ orbitals]
and $| l m_z \rangle$ is the eigenstate of the $z$ component $l_z$
of the orbital angular momentum operator with the eigenvalue $m_z$.

In these basis states,
the matrix elements of $l_x$ and $l_y$ are zero.
Then, we can evaluate the matrix elements of
the SOC term $\lambda \mib{l} \cdot \mib{s}$ easily
and its second-quantized form is given as follows:
\begin{equation}
  H_{\text{SOC}}
  =-i \frac{\lambda}{2} \sum_{\mib{r} \tau \sigma} \tau \sigma
  c^{\dagger}_{\mib{r} \tau \sigma} c_{\mib{r} \bar{\tau} \sigma},
  \label{eq:SOC}
\end{equation}
where $\lambda$ is the coupling constant of the SOC
and
$c^{\dagger}_{\mib{r} \tau \sigma}$ is the creation operator of the electron
with orbital $\tau$ ($=x$ or $y$)
and spin $\sigma$ ($=\uparrow$ or $\downarrow$) at site $\mib{r}$.
$\bar{\tau}=y$ ($x$) for $\tau=x$ ($y$).
We have also used the following notations:
$\tau=+1$ ($-1$) for the $x$ ($y$) orbital
and $\sigma=+1$ ($-1$) for up (down) spin.
We can assume $\lambda \ge 0$ without loss of generality
since the sign of $\lambda$ can be changed by the transformation
$c_{\mib{r} \tau \sigma} \rightarrow \tau c_{\mib{r} \tau \sigma}$
without changing the other terms of the model.

The kinetic energy term of the Hamiltonian is given by
\begin{equation}
  H_{\text{kin}}
  =
  \sum_{\mib{k},\tau,\sigma}
  \epsilon_{\mib{k} \tau}
  c^{\dagger}_{\mib{k} \tau \sigma}c_{\mib{k} \tau \sigma},
\end{equation}
where
$c^{\dagger}_{\mib{k} \tau \sigma}$ is the Fourier transform of
$c^{\dagger}_{\mib{r} \tau \sigma}$.
We consider only the nearest-neighbor hoppings
and the kinetic energies are given by
$\epsilon_{\mib{k} x} = -2 (t_1 \cos k_x+t_2 \cos k_y)$
and
$\epsilon_{\mib{k} y} = -2 (t_2 \cos k_x+t_1 \cos k_y)$.
We have set the lattice constant as unity.
The hopping integrals can be written in terms of
the two-center integrals~\cite{Slater1954}:
$t_1=(pp\sigma)$ and $t_2=(pp\pi)$ for the model for the $(p_x,p_y)$ orbitals
and
$t_1=(dd\pi)$ and $t_2=(dd\delta)$ for the model
for the $(d_{zx},d_{yz})$ orbitals.
We assume that both $t_1$ and $t_2$ are positive.
Then, the bandwidth is $W=8t$ with $t=(t_1+t_2)/2$.
In the presence of the SOC, the two bands split
but the bandwidth of each remains as $W=8t$,
irrespective of the value of $\lambda$.

The onsite interaction term is given by
\begin{equation}
  \begin{split}
    H_{I}&=
    U \sum_{\mib{r}, \tau}
    n_{\mib{r} \tau \uparrow} n_{\mib{r} \tau \downarrow}
    +U^{\prime} \sum_{\mib{r}}
    n_{\mib{r} x} n_{\mib{r} y}\\
    &+ J \sum_{\mib{r},\sigma,\sigma^{\prime}}
    c^{\dagger}_{\mib{r} x \sigma}
    c^{\dagger}_{\mib{r} y \sigma^{\prime}}
    c_{\mib{r} x \sigma^{\prime}}
    c_{\mib{r} y \sigma}\\
    &+J^{\prime}\sum_{\mib{r},\tau \ne \tau^{\prime}}
    c^{\dagger}_{\mib{r} \tau \uparrow}
    c^{\dagger}_{\mib{r} \tau \downarrow}
    c_{\mib{r} \tau^{\prime} \downarrow}
    c_{\mib{r} \tau^{\prime} \uparrow},
  \end{split}
\end{equation}
where 
$n_{\mib{r} \tau \sigma}=c^{\dagger}_{\mib{r} \tau \sigma}c_{\mib{r} \tau \sigma}$
and
$n_{\mib{r} \tau}=\sum_{\sigma}n_{\mib{r} \tau \sigma}$.
$U$ and $U'$ are the intraorbital and interorbital Coulomb interactions,
respectively.
$J$ is Hund's coupling and $J'$ denotes the pair-hopping interaction.
We use the relations $U=U'+J+J'$ and $J=J'$,
which hold in many orbitally degenerate systems
such as the $(p_x,p_y)$ and $(d_{zx},d_{yz})$ orbital systems~\cite{Tang1998}.

\section{Hartree-Fock Approximation for the Effective SOC}\label{sec:HF}
Before proceeding to the VMC calculation,
we discuss the effective SOC in the HF approximation.
In the presence of the SOC,
the expectation value of the operator
$c^{\dagger}_{\mib{r} \tau \sigma} c_{\mib{r} \bar{\tau} \sigma}$
connecting different orbitals becomes finite.
We denote it as
$\langle c^{\dagger}_{\mib{r} \tau \sigma} c_{\mib{r} \bar{\tau} \sigma} \rangle
= -i \tau \sigma K$.
Then, the following Fock term becomes finite
in the HF Hamiltonian:
\begin{equation}
  -i(U'-J')K
  \sum_{\mib{r} \tau \sigma} \tau \sigma
  c^{\dagger}_{\mib{r} \tau \sigma} c_{\mib{r} \bar{\tau} \sigma}.
  \label{eq:Fock}
\end{equation}
Note that the Fock term from Hund's coupling
disappears due to
$\sum_{\sigma}
\langle c^{\dagger}_{\mib{r} \tau \sigma} c_{\mib{r} \bar{\tau} \sigma}\rangle=0$.
The term~\eqref{eq:Fock} has the same form as the SOC term Eq.~\eqref{eq:SOC}
and can be combined into an effective SOC term.
The coupling constant of the effective SOC term is given by
\begin{equation}
  \lambda_{\text{eff}}=\lambda+2(U'-J')K.
\end{equation}
We call it as the effective SOC constant.
We determine $\lambda_{\text{eff}}$ self-consistently
by evaluating $K$ for a given $\lambda_{\text{eff}}$.
For physically reasonable parameters, that is,
$U'>J'$, $\lambda_{\text{eff}}$ becomes larger than the bare value $\lambda$.

For a weak $\lambda_{\text{eff}}$,
$K$ should be proportional to $\lambda_{\text{eff}}$
and we represent it as $K=\chi_{\text{mix}}\lambda_{\text{eff}}$.
In such a case, we obtain
\begin{equation}
  \lambda_{\text{eff}}=
  \frac{\lambda}{1 - 2(U'-J')\chi_{\text{mix}}}.
  \label{eq:HF}
\end{equation}
When the dispersions are the same for both orbitals,
that is,
$\epsilon_{\mib{k} x}=\epsilon_{\mib{k} y}$,
we can show $\chi_{\text{mix}}=\rho(\epsilon_{F})/2$,
where $\rho(\epsilon_{F})$ is the density of states at the Fermi level.
In the HF approximation,
$\lambda_{\text{eff}}$ depends on the interactions only through $(U'-J')$
and for a small $\lambda_{\text{eff}}$, it is proportional to $\lambda$.

At $2(U'-J')\chi_{\text{mix}}=1$, Eq.~\eqref{eq:HF} diverges.
However, the assumption to derive Eq.~\eqref{eq:HF}
does not hold there;
the nonlinear dependence of $K$ on $\lambda_{\text{eff}}$ is significant.
As a result,
$\lambda_{\text{eff}}$ remains finite even at $2(U'-J')\chi_{\text{mix}} \ge 1$
even within the HF approximation.

In Ref.~\citen{Liu2021},
an expression different from Eq.~\eqref{eq:HF} was obtained
for $\lambda_{\text{eff}}$
by using the eigenstates of $l_z$, $|l 1 \rangle$ and $|l -1 \rangle$,
as the basis states
instead of $|x \rangle$ and $|y \rangle$ [Eqs.~\eqref{eq:x} and ~\eqref{eq:y}].
For this basis, the interaction parameters are different
and here we denote them as $\tilde{U}$,  $\tilde{U}'$ and $\tilde{J}$.
In Ref.~\citen{Liu2021}, the combination of the interaction parameters
$(2\tilde{U}'-\tilde{U}-\tilde{J})$ appears instead of $(U'-J')$
in Eq.~\eqref{eq:HF}.
We can check the equivalence of them, for example for $l=2$, as follows.
For $l=2$, by using Racah parameters,
we obtain $\tilde{U}=\tilde{U}'=A+B+2C$
and $\tilde{J}=6B+2C$~\cite{Griffith1961}
and then, $(2\tilde{U}'-\tilde{U}-\tilde{J})=A-5B$.
For our expression for $l=2$, $U'=A-2B+C$ and $J'=3B+C$~\cite{Tang1998},
and we find $(U'-J')=A-5B=(2\tilde{U}'-\tilde{U}-\tilde{J})$.
Thus, the expression obtained in Ref.~\citen{Liu2021} is equivalent to ours.
However, we note that $\tilde{U} \ne \tilde{U}'+2\tilde{J}$,
while we can use $U=U'+2J$ and $J=J'$ in our basis.
A similar discussion to Ref.~\citen{Liu2021} is found in Ref.~\citen{Liu2008}.

\section{Variational Wave Function}
The DHWF is given by
\begin{equation}
  | \Psi_{\text{DHWF}} \rangle
  = P_d P_h P_{G} | \Phi \rangle,
\end{equation}
where $| \Phi \rangle$ is a one-body wave function
and $P_d$, $P_h$, and $P_G$ describe electron correlations.
This wave function is an adequate variational wave function
for the two-orbital Hubbard model at least for $\lambda=0$
around quarter filling, that is,
the electron number per site $n=1$~\cite{Kubo2021}.

The Gutzwiller projection operator $P_G$
describing the onsite correlations
is defined as~\cite{Okabe1997,Bunemann1998,Kobayashi2006,Kubo2009JPCS,
  Kubo2009PRB,Kubo2011JPSJSA,Kubo2017JPSJ,Kubo2021}
\begin{equation}
  P_{G}
  =\prod_{\mib{r} \gamma}
  \left[ 1-(1-g_{\gamma})P_{\mib{r}\gamma} \right],
\end{equation}
where $\gamma$ denotes one of the 16 onsite states,
$P_{\mib{r} \gamma}$ is the projection operator
onto state $\gamma$ at site $\mib{r}$,
and $g_{\gamma}$ is a variational parameter.
We assign $\gamma=0$ to the holon state, i.e., empty state.
By using the conservation of the number of electrons for each spin
and the equivalence of the orbital states,
we can reduce the number of $g_{\gamma}$ to be optimized to 5
in the paramagnetic (PM) and antiferromagnetic (AFM) states.

When the onsite Coulomb interactions, $U$ and $U'$, are strong and $n \simeq 1$,
most sites are occupied by a single electron.
In this situation,
if a doubly occupied site (doublon) is created,
an empty site (holon) should be around it
to reduce the energy by using singly-occupied virtual-states.
$P_d$ and $P_h$ describe such doublon-holon binding effects.
$P_d$ is an operator to include intersite correlation effects
concerning the doublon states.
This is defined as follows for the two-orbital model:~\cite{Kubo2021}
\begin{equation}
  P_d = \prod_{\mib{r}\, \gamma \in D}
  \left[1
    -(1-\zeta_{\gamma})P_{\mib{r} \gamma}
    \prod_{\mib{a}}(1-P_{\mib{r}+\mib{a} 0}) \right],
  \label{P_doublon}
\end{equation}
where $D$ denotes the set of doublon states,
i.e., onsite states with two electrons,
and $\mib{a}$ denotes the vectors connecting the nearest-neighbor sites.
$P_d$ gives factor $\zeta_{\gamma}$
when site $\mib{r}$ is in doublon state $\gamma$
and there is no holon at nearest-neighbor sites $\mib{r}+\mib{a}$.
Similarly,
$P_h$ describing the intersite correlation effects
on the holon state is defined as
\begin{equation}
  P_h = \prod_{\mib{r}}
  \left[1
    -(1-\zeta_{0}) P_{\mib{r} 0}
    \prod_{\mib{a}\, \gamma\in D}(1-P_{\mib{r}+\mib{a} \gamma}) \right].
  \label{P_holon}
\end{equation}
Factor $\zeta_0$ appears
when a holon exists without a nearest-neighboring doublon.
Considering symmetry, four $\zeta_{\gamma}$ are independent
variational parameters in the PM and AFM states.

For the one-body part $|\Phi\rangle$ of the wave function,
we consider an effective Hamiltonian:
\begin{equation}
  H^{\text{(eff)}}_{\mib{k} \sigma} =
  \begin{pmatrix}
    \epsilon_{\mib{k} x} & -i \sigma \lambda_{\text{eff}}/2
    & -\Delta_{\mib{Q} x \sigma} & -i \lambda_{\mib{Q}}/2 \\
    i \sigma \lambda_{\text{eff}}/2 & \epsilon_{\mib{k} y}
    & i \lambda_{\mib{Q}}/2 & -\Delta_{\mib{Q} y \sigma} \\
    -\Delta_{\mib{Q} x \sigma} & -i \lambda_{\mib{Q}}/2
    & \epsilon_{\mib{k}+\mib{Q} x} & -i \sigma \lambda_{\text{eff}}/2 \\
    i \lambda_{\mib{Q}}/2 & -\Delta_{\mib{Q} y \sigma}
    & i \sigma \lambda_{\text{eff}}/2 & \epsilon_{\mib{k}+\mib{Q} y}
  \end{pmatrix},
  \label{Heff}
\end{equation}
where
$\mib{Q}=(\pi,\pi)$ is the ordering vector and
$\Delta_{\mib{Q} \tau \sigma}
= \sigma \Delta_{s \, \mib{Q}} + \tau \Delta_{o \, \mib{Q}}$.
For an AFM ordered state,
$\Delta_{s \, \mib{Q}}$ becomes finite
and for an antiferro-orbital ordered state,
$\Delta_{o \, \mib{Q}}$ becomes finite.
We consider the effective SOC $\lambda_{\text{eff}}$
since we had to replace the bare SOC constant $\lambda$
by an effective one even in the HF approximation.
$\lambda_{\mib{Q}}$ denotes the spin- and site-dependent part
of the effective SOC in an AFM ordered state.
The term proportional to $\lambda_{\mib{Q}}$ can also be regarded as
a spin-independent site-dependent orbital mixing term.
$\Delta_{s \, \mib{Q}}$, $\Delta_{o \, \mib{Q}}$, $\lambda_{\mib{Q}}$,
and $\lambda_{\text{eff}}$ are variational parameters.
Similar variational parameters have been used
for the periodic Anderson model~\cite{Kubo2015JPCS,Kubo2015JPSJ,Kubo2015PP}.
In addition, it is possible to treat $t_1$ and $t_2$ in $\epsilon_{\mib{k} \tau}$
in Eq.~\eqref{Heff} as variational parameters.
We do not consider such a band renormalization effect on $\epsilon_{\mib{k} \tau}$
here although it may be an important future problem.
We construct $|\Phi\rangle$ by filling electrons
from the bottom of the energy of this effective Hamiltonian.

It is known that, at least for $\lambda=0$,
the stabilization of a partially spin-polarized ferromagnetic state
is difficult~\cite{Momoi1998,Sakamoto2002,
  Kubo2009JPCS,Kubo2009PRB,DeFranco2018,Maurya2022}.
Thus, we consider the completely polarized state
only with the majority-spin electrons as the ferromagnetic (FM) state.
The numbers of $g_{\gamma}$ and $\zeta_{\gamma}$ to be optimized are
reduced to 1 and 2, respectively, in the FM state
since the model is equivalent to a single-orbital Hubbard model
(with inter-spin mixing).
In the FM state, we consider antiferro-orbital order.

We optimize the variational parameters in the wave function
to reduce the expectation value of the energy
evaluated by the Monte Carlo method.
The momentum distribution function is also calculated
using the Monte Carlo method
for the optimized variational parameters.
If we set $g_{\gamma}=1$ and $\zeta_{\gamma}=1$ for all $\gamma$
and optimize only the variational parameters in the one-body part,
we obtain the HF results.

\section{Results}
In the following, we show results for an $L \times L$ square lattice
with antiperiodic-periodic boundary conditions with $L=12$.
To examine the finite-size effect, we also show some results for $L=8$ and $10$.
The number of electrons per site is fixed as $n=1$, i.e., quarter filling,
unless otherwise stated.

First, we discuss the finite-size effect on $\lambda_{\text{eff}}$.
\begin{figure}
  \includegraphics[width=0.99\linewidth]
    {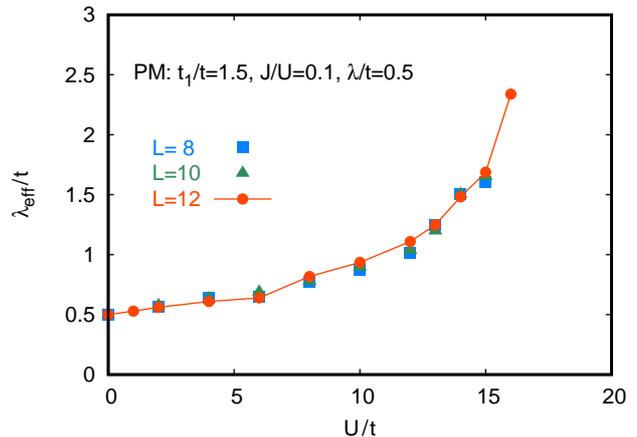}%
  \caption{(Color online)
    Effective SOC $\lambda_{\text{eff}}$ as a function of $U$
    for $t_1/t=1.5$, $J/U=0.1$, and $\lambda/t=0.5$
    for $L=8$ (squares), $L=10$ (triangles), and $L=12$ (circles).
    \label{lambda_size_dep}}
  \vspace{-4mm}
\end{figure}
In Fig.~\ref{lambda_size_dep},
we show $\lambda_{\text{eff}}$ as a function of $U$
for $t_1/t=1.5$, $J/U=0.1$, and $\lambda/t=0.5$
as an example for $L=8$, $10$, and $12$.
The finite-size effect on $\lambda_{\text{eff}}$ is weak.
For $U/t \gtrsim 16$, this solution becomes unstable.
We will discuss this point later.

\begin{figure}
  \includegraphics[width=0.99\linewidth]
    {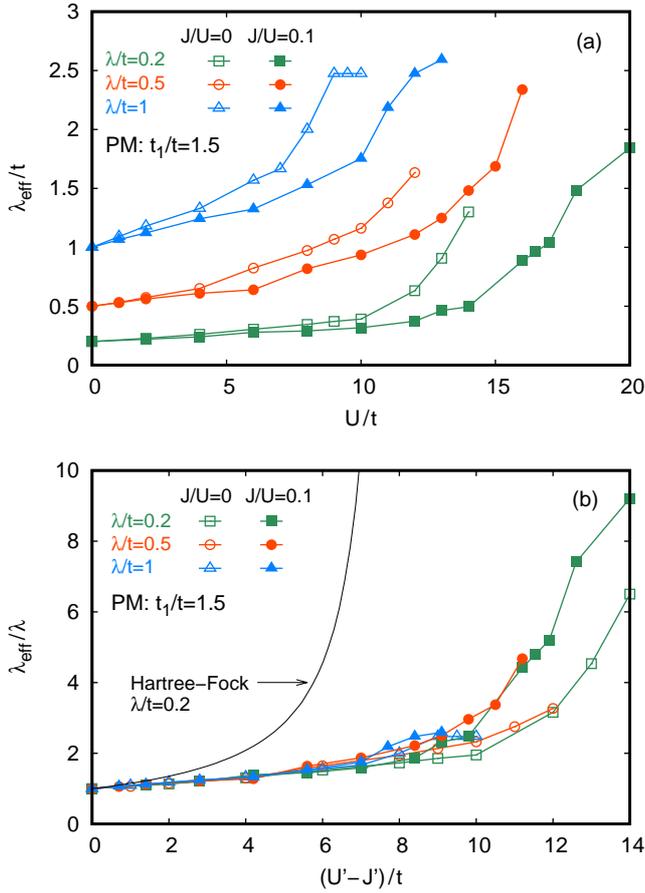}
  \caption{(Color online)
    Effective SOC $\lambda_{\text{eff}}$ for $t_1/t=1.5$
    for $J/U=0$ (open symbols) and $J/U=0.1$ (solid symbols) 
    for $\lambda/t=0.2$ (squares), $\lambda/t=0.5$ (circles),
    and $\lambda/t=1$ (triangles).
    (a) $\lambda_{\text{eff}}$ as a function of $U$.
    (b) $\lambda_{\text{eff}}$ normalized by the bare value $\lambda$
    as a function of $(U'-J')$.
    The solid line is the result of the HF approximation
    for $\lambda/t=0.2$.
    \label{lambda_vs_U_UpmJp}}
\end{figure}

Figure~\ref{lambda_vs_U_UpmJp}(a) shows
$\lambda_{\text{eff}}$ as a function of $U$
for $\lambda/t=0.2$, $0.5$, and $1$
with and without Hund's coupling.
$\lambda_{\text{eff}}$ increases as $U$ in all cases.
From the result of the HF approximation obtained in Sect.~\ref{sec:HF},
we expect that $\lambda_{\text{eff}}$ depends on the interactions
through $(U'-J')$ and is proportional to the bare value $\lambda$
at least for a small value of $\lambda_{\text{eff}}$.
Then, we plot $\lambda_{\text{eff}}/\lambda$
as a function of $(U'-J')$ in Fig.~\ref{lambda_vs_U_UpmJp}(b).
The data almost collapse on a single line for $(U'-J')/t \lesssim 6$.
For comparison, we draw the result of the HF approximation
for $\lambda/t=0.2$.
In the HF approximation, $\lambda_{\text{eff}}$
depends on the interactions only through $(U'-J')$ and
we numerically find that
$\lambda_{\text{eff}}/\lambda$ depends weakly on $\lambda$
for $\lambda_{\text{eff}} \lesssim \lambda_c$.
Here, $\lambda_c$ is defined as follows:
the upper band becomes empty for $\lambda>\lambda_c$
in the non-interacting case.
$\lambda_c/t=3.43$ for $t_1/t=1.5$ and $n=1$.
In the weak coupling region,
where the HF approximation is valid
and the values obtained with the VMC method
are near to those of the HF approximation,
the data collapse is obvious
since this scaling holds in the HF approximation.
However, the data collapse occurs in a relatively wide region;
the weak coupling region seems to be $(U'-J')/t \lesssim 1$
from Fig.~\ref{lambda_vs_U_UpmJp}(b).
For the large electron-electron interactions,
$\lambda_{\text{eff}}$ enhances several times larger than the bare value
although being strongly suppressed in comparison with the HF approximation.

\begin{figure}
  \includegraphics[width=0.99\linewidth]
    {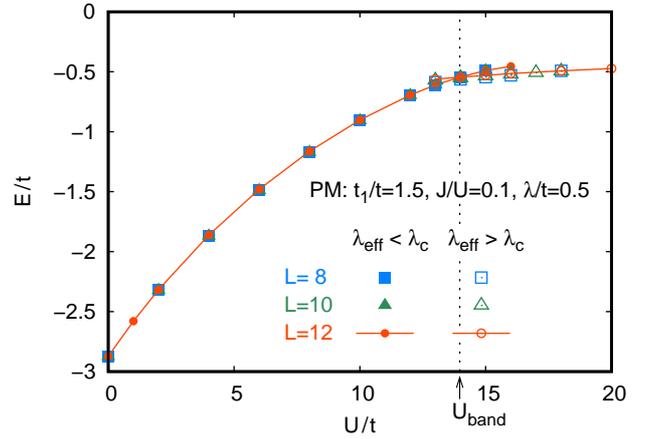}
  \caption{(Color online)
    Energy $E$ per site as a function of $U$
    for $t_1/t=1.5$, $J/U=0.1$, and $\lambda/t=0.5$
    for $L=8$ (squares), $L=10$ (triangles), and $L=12$ (circles)
    in the PM phase.
    There are two solutions
    with $\lambda_{\text{eff}}<\lambda_c$ (solid symbols)
    and $\lambda_{\text{eff}}>\lambda_c$ (open symbols).
    $\lambda_c/t=3.43$ for $t_1/t=1.5$.
    The vertical dotted line indicates
    the first-order band-structure transition point $U_{\text{band}}$
    for $L=12$.
    \label{E_size_dep}}
\end{figure}

\begin{figure}
  \includegraphics[width=0.99\linewidth]
    {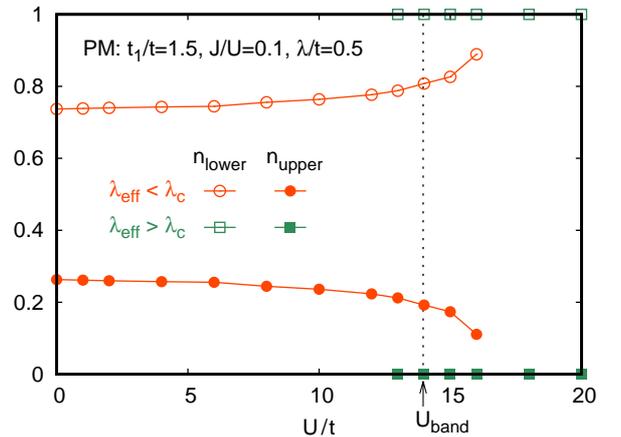}
  \caption{(Color online)
    Electron occupation numbers per site
    of the lower band $n_{\text{lower}}$ (open symbols)
    and of the upper band $n_{\text{upper}}$ (solid symbols)
    as functions of $U$
    for $\lambda_{\text{eff}}<\lambda_c$ (circles)
    and for $\lambda_{\text{eff}}>\lambda_c$ (squares)
    for $t_1/t=1.5$, $J/U=0.1$, and $\lambda/t=0.5$.
    The vertical dotted line indicates
    the first-order band-structure transition point $U_{\text{band}}$.
    \label{filling}}
\end{figure}

Figure~\ref{E_size_dep} shows energy as a function of $U$
for $t_1/t=1.5$, $J/U=0.1$, and $\lambda/t=0.5$ in the PM phase.
We obtain two solutions in the PM phase:
one is with $\lambda_{\text{eff}}<\lambda_c$
and the other is with $\lambda_{\text{eff}}>\lambda_c$.
For a large $U$,
the solution with $\lambda_{\text{eff}}<\lambda_c$ becomes unstable
and the solution with $\lambda_{\text{eff}}>\lambda_c$ appears.
In Figs.~\ref{lambda_size_dep} and \ref{lambda_vs_U_UpmJp},
we have shown the results with $\lambda_{\text{eff}}<\lambda_c$.
In the solution with $\lambda_{\text{eff}}>\lambda_c$,
the upper band of the effective Hamiltonian is empty
as shown in Fig.~\ref{filling}
and thus, this solution can be regarded as a one-band state.
There is a first-order band-structure transition
from the two-band state to the one-band state
at $U=U_{\text{band}} \simeq 14t$ for this parameter set.
We also notice that the lattice size dependence on the energy is weak.

\begin{figure}
  \includegraphics[width=0.99\linewidth]
    {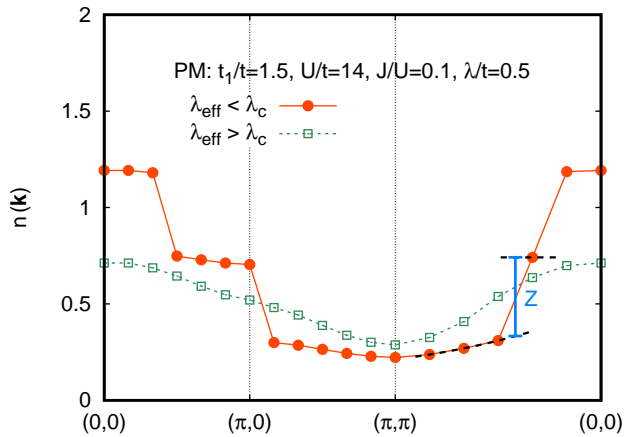}
  \caption{(Color online)
    Momentum distribution function
    $n(\mib{k})=\sum_{\tau}
    \langle c^{\dagger}_{\mib{k} \tau \sigma} c_{\mib{k} \tau \sigma} \rangle$
    for $t_1/t=1.5$, $U/t=14$, $J/U=0.1$, and $\lambda/t=0.5$
    with $\lambda_{\text{eff}}<\lambda_c$ (circles)
    and with $\lambda_{\text{eff}}>\lambda_c$ (squares).
    The renormalization factor $Z$ is estimated
    by extrapolating $n(\mib{k})$
    from above and below the Fermi momentum
    along $(\pi,\pi)$--$(0,0)$
    (dashed lines; demonstrated for $\lambda_{\text{eff}}<\lambda_c$).
    Due to the antiperiodic boundary condition for the $x$ direction,
    we shift $k_x$ by $\pi/L$ ($L=12$); for example,
    $(\pi,\pi)$ denoted in this figure actually means
    the point $(\pi-\pi/L,\pi)$.
    \label{nk}}
\end{figure}

The first-order band-structure transition for this parameter set
accompanies the Mott insulator transition.
To show it we discuss the momentum distribution function
$n(\mib{k})=\sum_{\tau}
\langle c^{\dagger}_{\mib{k} \tau \sigma} c_{\mib{k} \tau \sigma} \rangle$.
In Fig.~\ref{nk}, we show $n(\mib{k})$
for $\lambda_{\text{eff}}<\lambda_c$
and $\lambda_{\text{eff}}>\lambda_c$
at $U/t=14$.
For $\lambda_{\text{eff}}<\lambda_c$, $n(\mib{k})$ jumps at the Fermi momenta,
that is, it is in a metallic state.
On the other hand, Fermi surface disappears for $\lambda_{\text{eff}}>\lambda_c$,
that is, it is in an insulating state.
For a quantitative discussion,
we evaluate the renormalization factor $Z$
by the first jump along $(\pi,\pi)$--$(0,0)$,
as demonstrated for $\lambda_{\text{eff}}<\lambda_c$ in Fig~\ref{nk}.
The renormalization factor is inversely proportional to the effective mass
and becomes zero in the Mott insulating state.

\begin{figure}
  \includegraphics[width=0.99\linewidth]
    {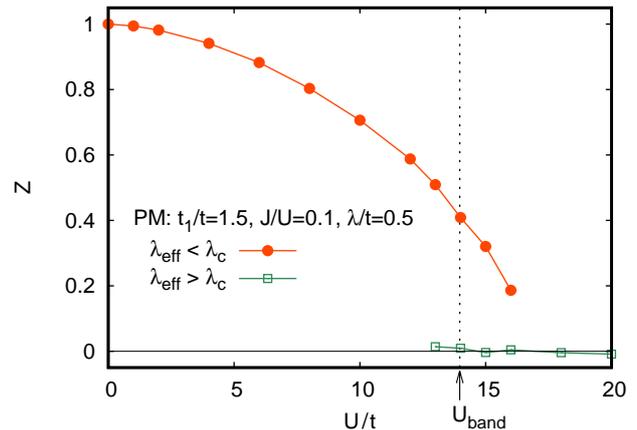}
  \caption{(Color online)
    Renormalization factor $Z$ as a function of $U$
    for $t_1/t=1.5$, $J/U=0.1$, and $\lambda/t=0.5$
    with $\lambda_{\text{eff}} < \lambda_c$ (solid circles)
    and  $\lambda_{\text{eff}} > \lambda_c$ (open circles).
    The vertical dotted line indicates
    the first-order band-structure transition point $U_{\text{band}}$.
    \label{Z}}
\end{figure}

In Fig.~\ref{Z}, we show $Z$ as a function of $U$
for $t_1/t=1.5$, $J/U=0.1$, and $\lambda/t=0.5$.
$Z$ remains finite for the solution with $\lambda < \lambda_c$
and becomes zero for the solution with $\lambda > \lambda_c$.
Thus, for this parameter set,
the first-order band-structure transition at $U_{\text{band}}$ accompanies
the Mott metal-insulator transition.

\begin{figure}
  \includegraphics[width=0.99\linewidth]
    {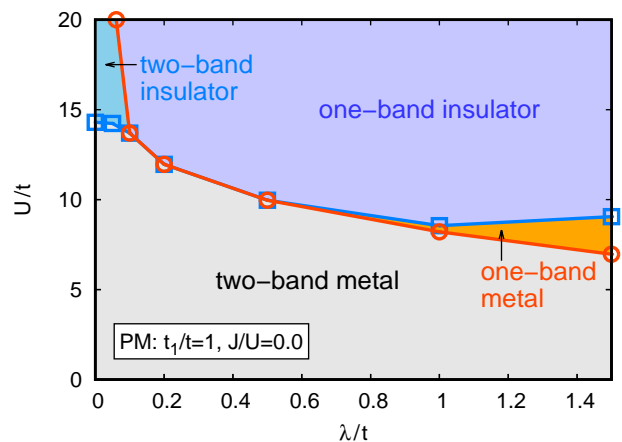}
  \caption{(Color online)
    Phase diagram for $t_1/t=1$ and $J/U=0$ in the PM phase.
    The circles represent the band-structure transition points $U_{\text{band}}$.
    The squares represent
    the Mott metal-insulator transition points $U_{\text{MIT}}$.
    \label{PD_PM_t11_J.0U}}
\end{figure}

\begin{figure}
  \includegraphics[width=0.99\linewidth]
    {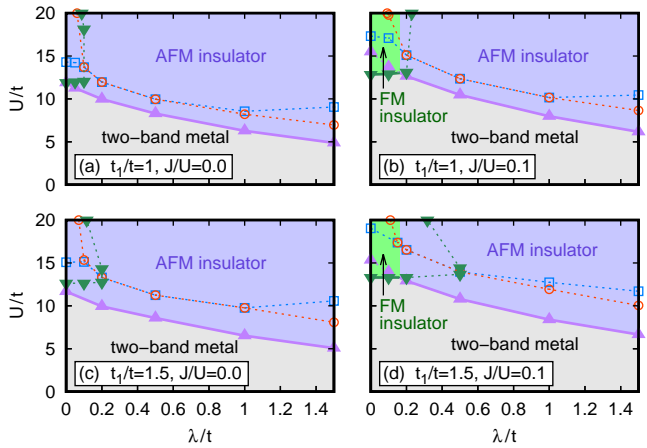}
  \caption{(Color online)
    Magnetic phase diagrams
    (a) for $t_1/t=1$ and $J/U=0$,
    (b) for $t_1/t=1$ and $J/U=0.1$,
    (c) for $t_1/t=1.5$ and $J/U=0$, and
    (d) for $t_1/t=1.5$ and $J/U=0.1$.
    The circles represent the band-structure transition points $U_{\text{band}}$
    assuming the PM phase.
    The squares denote
    the Mott metal-insulator transition points $U_{\text{MIT}}$
    assuming the PM phase.
    The triangles indicate the AFM transition points $U_{\text{AFM}}$
    from the PM phase.
    The downward triangles represent the FM transition points $U_{\text{FM}}$
    from the PM phase.
    \label{PD}}
\end{figure}

In Fig.~\ref{PD_PM_t11_J.0U}, we show a phase diagram
for $t_1/t=1$ and $J/U=0$ in the PM phase.
As shown in Figs.~\ref{E_size_dep} and \ref{filling},
the band-structure transition from the two-band state to the one-band state
can occur by increasing $U$.
In addition, Mott transitions are possible to occur.
In a finite-size lattice with $L \times L$,
it may be difficult to estimate $Z$ smaller than $1/L$ (see Fig.~\ref{nk}).
Thus, we determine the Mott metal-insulator transition point $U_{\text{MIT}}$
by extrapolating data with $Z \gtrsim 0.1$ to zero as a function of $U$.
Then, we obtain four phases in the PM phase:
two-band metal, two-band insulator, one-band metal, and one-band insulator.
Without the SOC, the model is an ordinary two-orbital Hubbard model,
and we find the Mott transition
in the two-band state~\cite{Bunemann1998,Han1998,Ono2003,Koga2006,
  deMedici2011PRB,deMedici2011PRL,Takenaka2012,Facio2017,DeFranco2018,Kubo2021}.
Thus, there are two-band metallic phase and two-band insulating phase
for $\lambda=0$.
For a sufficiently large value of $\lambda$,
the band-structure transition from the two-band state to the one-band state
easily occurs even by a small Coulomb interaction.
Thus, a transition to the one-band metallic state occurs first,
and then, the Mott transition in the one-band state occurs by increasing $U$
as in the ordinary single-orbital Hubbard model~\cite{Yokoyama2002,Capello2006}.
In between, for example, $\lambda/t=0.2$ and $0.5$,
the Coulomb interaction is already larger
than $U_{\text{MIT}}$ of the single-orbital Hubbard model
when the band-structure transition to the one-band state occurs.
Thus, for these cases, the Mott transition simultaneously occurs
with the band-structure transition.
We obtain similar phase diagrams in the PM phase
for $t_1 \ne t_2$ and for a finite Hund's coupling
(see $U_{\text{band}}$ and $U_{\text{MIT}}$ in Fig.~\ref{PD}).

Figure~\ref{PD} shows magnetic phase diagrams
constructed by comparing the energies of the FM, AFM, and PM states.
In these phase diagrams,
we also show $U_{\text{band}}$ and $U_{\text{MIT}}$
assuming the PM phase for a comparison
while these transitions are masked by the magnetic order.
In the one-band phase above $U_{\text{band}}$,
the system should be in the AFM phase
since the model is effectively reduced
to the half-filled single-orbital Hubbard model,
in which the ground state is expected to be the AFM state for $U>0$
due to the perfect nesting of the Fermi surface.
The combination of the energy gain of
the AFM order and the band-structure transition
leads to the AFM transition at $U_{\text{AFM}} < U_{\text{band}}$
as shown in Fig.~\ref{PD}.
Here, $U_{\text{AFM}}$ is determined
by comparing the energies of the PM and AFM states.
Some details of the phase diagrams depend on $t_1 (=2t-t_2)$ and $J$.
First, we discuss the case of $t_1=t_2$ and $J=0$ shown in Fig.~\ref{PD}(a).
For $t_1=t_2$ and $J=0$, the orbital and spin degrees of freedom
are equivalent without the SOC.
In this case, the AFM order in the one-band state
is equivalent to the antiferro-orbital order in the FM state
only with the majority spin band.
Then, we obtain $U_{\text{AFM}}=U_{\text{FM}}$ within the numerical accuracy
for $\lambda=0$.
Here, the FM transition point $U_{\text{FM}}$ is determined
by comparing the energies of the PM and FM states.
For a finite $\lambda$, band splitting occurs
and the antiferro-orbital order supporting the FM state becomes unstable.
As a result, we obtain $U_{\text{FM}}>U_{\text{AFM}}$ for $\lambda>0$,
that is, the FM phase is limited to $\lambda=0$.
For a finite Hund's coupling $J$ shown in Fig.~\ref{PD}(b) with $t_1=t_2$,
the orbital and spin are not equivalent even for $\lambda=0$.
At $\lambda=0$, the model is the ordinary two-orbital Hubbard model
and the FM transition occurs first by increasing
$U$.~\cite{Kubo2009JPCS,Kubo2009PRB,DeFranco2018,Kubo2021}
We find that this FM phase extends for finite values of $\lambda$.
For $t_1 \ne t_2$, the orbital and spin are not equivalent
even for $J=0$ and $\lambda=0$.
For $t_1 \ne t_2$ and $J=0$ shown in Fig.~\ref{PD}(c),
we find that $U_{\text{FM}}$ is always larger than $U_{\text{AFM}}$.
For a finite $J$ and $t_1 \ne t_2$, we obtain a phase diagram
shown in Fig.~\ref{PD}(d).
The FM phase extends for finite values of $\lambda$ as in Fig.~\ref{PD}(b).
These results suggest that
Hund's coupling is important in realizing the FM phase.
We have not find a transition from the FM state to the AFM state
by increasing $U$,
that is, the energy of the FM state is always lower than that of the AFM state
when $U_{\text{FM}} < U_{\text{AFM}}$ even at $U>U_{\text{AFM}}$.
A transition from the AFM state to the FM state
is not realized, either, within the parameters we have searched.

Finally, we discuss the doping dependence of $\lambda_{\text{eff}}$.
For sufficiently doped cases,
we expect that the system remains
in the PM metallic phase~\cite{Kubo2009PRB,DeFranco2018}.
Thus, in the following, we consider only the PM states.

\begin{figure}
  \includegraphics[width=0.99\linewidth]
    {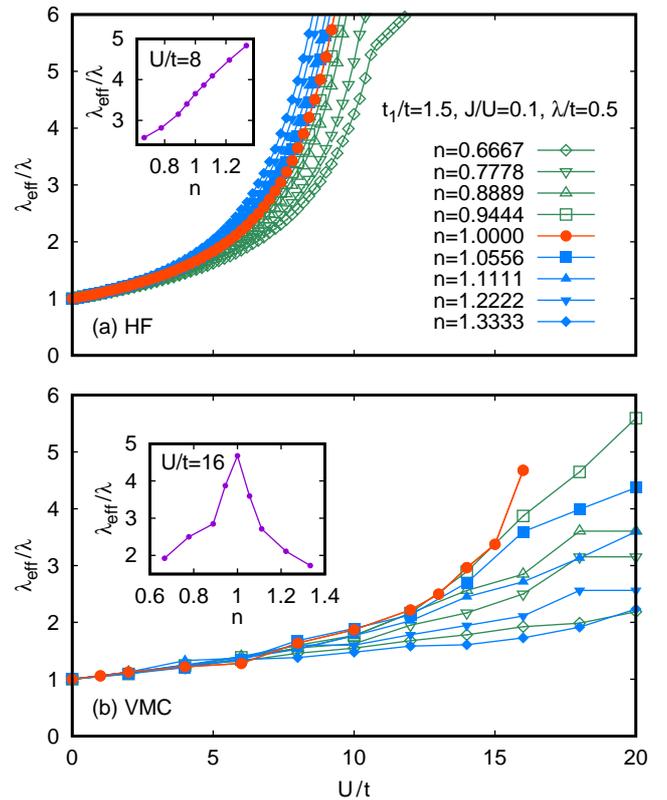}
  \caption{(Color online)
      Effective SOC constant $\lambda_{\text{eff}}$ as a function of $U$
      for $t_1/t=1.5$, $J/U=0.1$, and $\lambda/t=0.5$
      for various values of $n$.
      (a) Results obtained with the HF approximation.
      Inset shows $\lambda_{\text{eff}}$ as a function of $n$ for $U/t=8$.
      (b) Results obtained with the VMC method.
      Inset shows $\lambda_{\text{eff}}$ as a function of $n$ for $U/t=16$.
    \label{lambda_doped}}
\end{figure}
Figure \ref{lambda_doped}(a) shows $\lambda_{\text{eff}}$
for $t_1/t=1.5$, $J/U=0.1$, and $\lambda/t=0.5$
for various values of filling $n$ obtained with the HF approximation.
The upper band becomes empty when $\lambda_{\text{eff}}$ reaches $\lambda_c$.
In this figure, we see such band-structure transitions
for $n=0.6667$ ($\lambda_c/\lambda=5.40$)
and for $n=0.7778$ ($\lambda_c/\lambda=5.90$).
In the HF approximation,
$\lambda_{\text{eff}}$ varies monotonically with $n$ across $n=1$.
For a small value of $\lambda_{\text{eff}}$,
it depends on $n$ through $\chi_{\text{mix}}$
as shown in Eq.~\eqref{eq:HF}.
$\chi_{\text{mix}}$ is a smooth function of $n$ around $n=1$
since the band dispersion does not have a particular feature around there.
As a result, $\lambda_{\text{eff}}$ changes smoothly with $n$ across $n=1$.
This feature can be clearly observed by plotting $\lambda_{\text{eff}}$
as a function of $n$ for a fixed value of $U$ as shown in the inset.

On the other hand, $\lambda_{\text{eff}}$ obtained with the VMC method
enhances around the integer filling $n=1$ [Fig.~\ref{lambda_doped}(b)].
By doping from $n=1$, $\lambda_{\text{eff}}$ decreases.
It is also recognized by observing the $n$ dependence of $\lambda_{\text{eff}}$
at a fixed value of $U$ as in the inset of Fig.~\ref{lambda_doped}(b).
This result implies that $\lambda_{\text{eff}}$ enhances
around the Mott insulating phase
and it is important to include
the correlation effects beyond the HF approximation
to discuss the enhancement of the effective SOC properly.

\begin{figure}
  \includegraphics[width=0.99\linewidth]
    {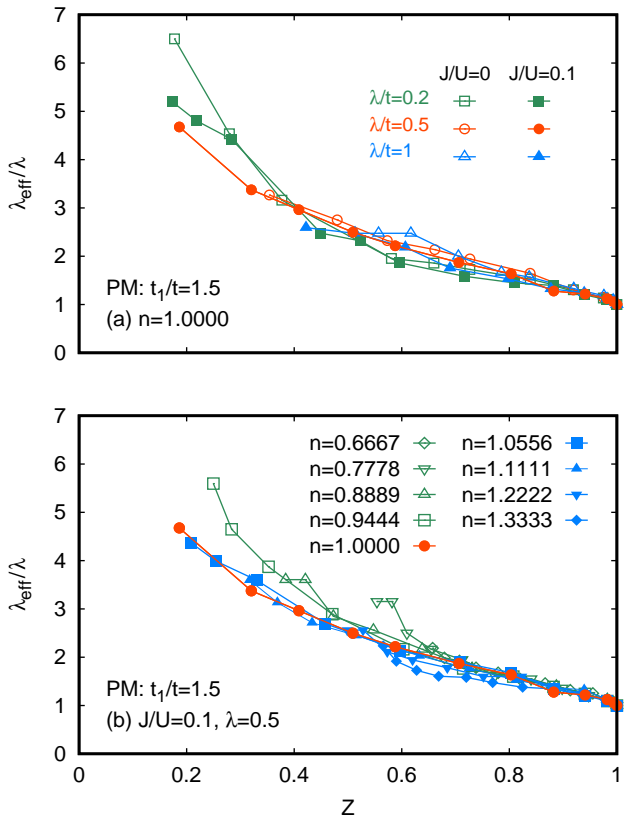}
  \caption{(Color online)
      Effective SOC constant $\lambda_{\text{eff}}$ as a function of $Z$
      for $t_1/t=1.5$.
      (a) For $n=1$
      for $J/U=0$ (open symbols) and $J/U=0.1$ (solid symbols) 
      for $\lambda/t=0.2$ (squares), $\lambda/t=0.5$ (circles),
      and $\lambda/t=1$ (triangles).
      (b) For various values of $n$ for $J/U=0.1$ and $\lambda/t=0.5$.
    \label{lambda_vs_Z}}
\end{figure}

To confirm these expectations,
we show $\lambda_{\text{eff}}$ as a function of $Z$ for $n=1$
with several combinations of $J$ and $\lambda$ [Fig.~\ref{lambda_vs_Z}(a)]
and for various values of $n$ with $J/U=0.1$ and $\lambda=0.5$
[Fig.~\ref{lambda_vs_Z}(b)].
In both cases, we find that $\lambda_{\text{eff}}$ increases as $Z$ decreases.
In other words,
$\lambda_{\text{eff}}$ enhances in the strongly correlated region
around the Mott transition point $Z=0$.

\section{Summary}
We have investigated the two-orbital Hubbard model
with the SOC on a square lattice
by applying the VMC method.
The effective SOC constant $\lambda_{\text{eff}}$ is introduced
as a variational parameter.
We have mainly investigated the case of $n=1$.

In the PM metallic state,
$\lambda_{\text{eff}}$ depends on the interactions through $(U'-J')$
and is proportional to the bare value $\lambda$
in a relatively wide parameter region.
This behavior is expected within the HF approximation,
but we find this behavior up to $(U'-J')/t \simeq 6$,
where the HF approximation is already inadequate.
We also find that $\lambda_{\text{eff}}$ can become several times larger
than the bare value by the Coulomb interaction.

Within the PM phase,
we find the band-structure transition from a two-band state to a one-band state
by increasing the Coulomb interaction due to the enhanced effective SOC.
Then, we obtain four phases in the PM phase:
two-band metallic, two-band Mott insulating,
one-band metallic, and one-band Mott insulating phases.

By considering magnetic order,
we find that AFM order in the effective one-band state occurs in a wide region
by combining energy gain by the band-structure transition and AFM order.
The FM insulating phase appears for a small $\lambda$
with a finite Hund's coupling.
This FM ordered state becomes unstable when $\lambda$ is increased
since the antiferro-orbital order supporting the FM state
is destabilized by the band splitting.

We have also investigated the carrier doping effects.
In the results of the VMC method,
$\lambda_{\text{eff}}$ is enhanced most at $n=1$.
It is in a sharp contrast to the HF approximation,
in which $\lambda_{\text{eff}}$ changes monotonically with $n$ across $n=1$.
Thus, it is necessary to include the correlation effects
beyond the HF approximation
to discuss the enhanced effective SOC properly.
By discussing $\lambda_{\text{eff}}$ for $n=1$ and doped cases
as a function of the renormalization factor $Z$,
we conclude that $\lambda_{\text{eff}}$ enhances
in the strongly correlated region around the Mott transition point.
Therefore, it should be interesting to search for SOC-originating phenomena
in strongly correlated metals in the vicinity
of the Mott insulating phase.


\begin{thebibliography}{10}

\bibitem{Dyakonov1971JETPL}
M.~I. Dyakonov and V.~I. Perel, JETP Lett. {\bfseries 13},  467 (1971).

\bibitem{Dyakonov1971PLA}
M.~I. Dyakonov and V.~I. Perel, Phys. Lett. {\bfseries 35A},  459 (1971).

\bibitem{Hirsch1999}
J.~E. Hirsch, Phys. Rev. Lett. {\bfseries 83},  1834 (1999).

\bibitem{Murakami2003}
S.~Murakami, N.~Nagaosa, and S.-C. Zhang, Science {\bfseries 301},  1348
  (2003).

\bibitem{Sinova2004}
J.~Sinova, D.~Culcer, Q.~Niu, N.~A. Sinitsyn, T.~Jungwirth, and A.~H.
  MacDonald, Phys. Rev. Lett. {\bfseries 92},  126603 (2004).

\bibitem{Kato2004}
Y.~K. Kato, R.~C. Myers, A.~C. Gossard, and D.~D. Awschalom, Science {\bfseries
  306},  1910 (2004).

\bibitem{Wunderlich2005}
J.~Wunderlich, B.~Kaestner, J.~Sinova, and T.~Jungwirth, Phys. Rev. Lett.
  {\bfseries 94},  047204 (2005).

\bibitem{Saitoh2006}
E.~Saitoh, M.~Ueda, H.~Miyajima, and G.~Tatara, Appl. Phys. Lett. {\bfseries
  88},  182509 (2006).

\bibitem{Valenzuela2006}
S.~O. Valenzuela and M.~Tinkham, Nature {\bfseries 442},  176 (2006).

\bibitem{Kimura2007}
T.~Kimura, Y.~Otani, T.~Sato, S.~Takahashi, and S.~Maekawa, Phys. Rev. Lett.
  {\bfseries 98},  156601 (2007).

\bibitem{Zutic2004}
I.~{\v Z}uti{\'c}, J.~Fabian, and S.~Das~Sarma, Rev. Mod. Phys. {\bfseries 76},
   323 (2004).

\bibitem{Kane2005PRL1}
C.~L. Kane and E.~J. Mele, Phys. Rev. Lett. {\bfseries 95},  226801 (2005).

\bibitem{Bernevig2006PRL}
B.~A. Bernevig and S.-C. Zhang, Phys. Rev. Lett. {\bfseries 96},  106802
  (2006).

\bibitem{Kane2005PRL2}
C.~L. Kane and E.~J. Mele, Phys. Rev. Lett. {\bfseries 95},  146802 (2005).

\bibitem{Bernevig2006Science}
B.~A. Bernevig, T.~L. Hughes, and S.-C. Zhang, Science {\bfseries 314},  1757
  (2006).

\bibitem{Konig2007}
M.~K{\"o}nig, S.~Wiedmann, C.~Br{\"u}ne, A.~Roth, H.~Buhmann, L.~W. Molenkamp,
  X.-L. Qi, and S.-C. Zhang, Science {\bfseries 318},  766 (2007).

\bibitem{Roth2009}
A.~Roth, C.~Br{\"u}ne, H.~Buhmann, L.~W. Molenkamp, J.~Maciejko, X.-L. Qi, and
  S.-C. Zhang, Science {\bfseries 325},  294 (2009).

\bibitem{Liu2008}
G.-Q. Liu, V.~N. Antonov, O.~Jepsen, and O.~K. Andersen., Phys. Rev. Lett.
  {\bfseries 101},  026408 (2008).

\bibitem{Veenstra2014}
C.~N. Veenstra, Z.-H. Zhu, M.~Raichle, B.~M. Ludbrook, A.~Nicolaou, B.~Slomski,
  G.~Landolt, S.~Kittaka, Y.~Maeno, J.~H. Dil, I.~S. Elfimov, M.~W. Haverkort,
  and A.~Damascelli, Phys. Rev. Lett. {\bfseries 112},  127002 (2014).

\bibitem{Zhang2016}
G.~Zhang, E.~Gorelov, E.~Sarvestani, and E.~Pavarini, Phys. Rev. Lett.
  {\bfseries 116},  106402 (2016).

\bibitem{Kim2018}
M.~Kim, J.~Mravlje, M.~Ferrero, O.~Parcollet, and A.~Georges, Phys. Rev. Lett.
  {\bfseries 120},  126401 (2018).

\bibitem{Tamai2019}
A.~Tamai, M.~Zingl, E.~Rozbicki, E.~Cappelli, S.~Ricc{\`o}, A.~{de la Torre},
  S.~McKeown~Walker, F.~Y. Bruno, P.~D.~C. King, W.~Meevasana, M.~Shi,
  M.~Radovi{\'c}, N.~C. Plumb, A.~S. Gibbs, A.~P. Mackenzie, C.~Berthod,
  H.~U.~R. Strand, M.~Kim, A.~Georges, and F.~Baumberger, Phys. Rev. X
  {\bfseries 9},  021048 (2019).

\bibitem{Linden2020}
N.-O. Linden, M.~Zingl, C.~Hubig, O.~Parcollet, and U.~Schollw{\"o}ck, Phys.
  Rev. B {\bfseries 101},  041101(R) (2020).

\bibitem{Cao2021}
X.~Cao, Y.~Lu, P.~Hansmann, and M.~W. Haverkort, Phys. Rev. B {\bfseries 104},
  115119 (2021).

\bibitem{Klejnberg1999}
A.~Klejnberg and J.~Spalek, J. Phys.: Condens. Matter {\bfseries 11},  6553
  (1999).

\bibitem{Han2004}
J.~E. Han, Phys. Rev. B {\bfseries 70},  054513 (2004).

\bibitem{Sakai2004}
S.~Sakai, R.~Arita, and H.~Aoki, Phys. Rev. B {\bfseries 70},  172504 (2004).

\bibitem{Kubo2007}
K.~Kubo, Phys. Rev. B {\bfseries 75},  224509 (2007).

\bibitem{Kubo2008JPSJ}
K.~Kubo, J. Phys. Soc. Jpn. {\bfseries 77},  043702 (2008).

\bibitem{Kubo2008JOAM}
K.~Kubo, J. Optoelectron. Adv. Mater. {\bfseries 10},  1683 (2008).

\bibitem{Cvetkovic2013}
V.~Cvetkovic and O.~Vafek, Phys. Rev. B {\bfseries 88},  134510 (2013).

\bibitem{Vafek2017}
O.~Vafek and A.~V. Chubukov, Phys. Rev. Lett. {\bfseries 118},  087003 (2017).

\bibitem{Yu2018}
Y.~Yu, A.~K.~C. Cheung, S.~Raghu, and D.~F. Agterberg, Phys. Rev. B {\bfseries
  98},  184507 (2018).

\bibitem{Cheung2019}
A.~K.~C. Cheung and D.~F. Agterberg, Phys. Rev. B {\bfseries 99},  024516
  (2019).

\bibitem{Ramires2019}
A.~Ramires and M.~Sigrist, Phys. Rev. B {\bfseries 100},  104501 (2019).

\bibitem{Kaba2019}
S.-O. Kaba and D.~S{\'e}n{\'e}chal, Phys. Rev. B {\bfseries 100},  214507
  (2019).

\bibitem{Huang2019}
W.~Huang, Y.~Zhou, and H.~Yao, Phys. Rev. B {\bfseries 100},  134506 (2019).

\bibitem{Suh2020}
H.~G. Suh, H.~Menke, P.~M.~R. Brydon, C.~Timm, A.~Ramires, and D.~F. Agterberg,
  Phys. Rev. Research {\bfseries 2},  032023(R) (2020).

\bibitem{Clepkens2021}
J.~Clepkens, A.~W. Lindquist, X.~Liu, and H.-Y. Kee, Phys. Rev. B {\bfseries
  104},  104512 (2021).

\bibitem{Kim2008}
B.~J. Kim, H.~Jin, S.~J. Moon, J.~Y. Kim, B.~G. Park, C.~S. Leem, J.~Yu, T.~W.
  Noh, C.~Kim, S.~J. Oh, J.~H. Park, V.~Durairaj, G.~Cao, and E.~Rotenberg,
  Phys. Rev. Lett. {\bfseries 101},  076402 (2008).

\bibitem{Moon2008}
S.~J. Moon, H.~Jin, K.~W. Kim, W.~S. Choi, Y.~S. Lee, J.~Yu, G.~Cao, A.~Sumi,
  H.~Funakubo, C.~Bernhard, and T.~W. Noh, Phys. Rev. Lett. {\bfseries 101},
  226402 (2008).

\bibitem{Kim2009}
B.~J. Kim, H.~Ohsumi, T.~Komesu, S.~Sakai, T.~Morita, H.~Takagi, and T.~Arima,
  Science {\bfseries 323},  1329 (2009).

\bibitem{Watanabe2010}
H.~Watanabe, T.~Shirakawa, and S.~Yunoki, Phys. Rev. Lett. {\bfseries 105},
  216410 (2010).

\bibitem{Wang2011}
F.~Wang and T.~Senthil, Phys. Rev. Lett. {\bfseries 106},  136402 (2011).

\bibitem{Qi2012}
T.~F. Qi, O.~B. Korneta, L.~Li, K.~Butrouna, V.~S. Cao, X.~Wan, P.~Schlottmann,
  R.~K. Kaul, and G.~Cao, Phys. Rev. B {\bfseries 86},  125105 (2012).

\bibitem{Watanabe2014}
H.~Watanabe, T.~Shirakawa, and S.~Yunoki, Phys. Rev. B {\bfseries 89},  165115
  (2014).

\bibitem{Triebl2018}
R.~Triebl, G.~J. Kraberger, J.~Mravlje, and M.~Aichhorn, Phys. Rev. B
  {\bfseries 98},  205128 (2018).

\bibitem{Richter2021}
M.~Richter, J.~Graspeuntner, T.~Sch{\"a}fer, N.~Wentzell, and M.~Aichhorn,
  Phys. Rev. B {\bfseries 104},  195107 (2021).

\bibitem{Bunemann2016}
J.~B{\"u}nemann, T.~Linneweber, U.~L{\"o}w, F.~B. Anders, and F.~Gebhard, Phys.
  Rev. B {\bfseries 94},  035116 (2016).

\bibitem{Liu2021}
Z.~Liu, J.-Y. You, B.~Gu, S.~Maekawa, and G.~Su, arXiv:2106.01046.

\bibitem{Yokoyama1987}
H.~Yokoyama and H.~Shiba, J. Phys. Soc. Jpn. {\bfseries 56},  1490 (1987).

\bibitem{Kaplan1982}
T.~A. Kaplan, P.~Horsch, and P.~Fulde, Phys. Rev. Lett. {\bfseries 49},  889
  (1982).

\bibitem{Yokoyama1990}
H.~Yokoyama and H.~Shiba, J. Phys. Soc. Jpn. {\bfseries 59},  3669 (1990).

\bibitem{Yokoyama2002}
H.~Yokoyama, Prog. Theor. Phys. {\bfseries 108},  59 (2002).

\bibitem{Capello2006}
M.~Capello, F.~Becca, S.~Yunoki, and S.~Sorella, Phys. Rev. B {\bfseries 73},
  245116 (2006).

\bibitem{Watanabe2006}
T.~Watanabe, H.~Yokoyama, Y.~Tanaka, and J.-i. Inoue, J. Phys. Soc. Jpn.
  {\bfseries 75},  074707 (2006).

\bibitem{Yokoyama2006}
H.~Yokoyama, M.~Ogata, and Y.~Tanaka, J. Phys. Soc. Jpn. {\bfseries 75},
  114706 (2006).

\bibitem{Onari2007}
S.~Onari, H.~Yokoyama, and Y.~Tanaka, Physica C {\bfseries 463--465},  120
  (2007).

\bibitem{Koga2006}
A.~Koga, N.~Kawakami, H.~Yokoyama, and K.~Kobayashi, AIP Conf. Proc. {\bfseries
  850},  1458 (2006).

\bibitem{Takenaka2012}
Y.~Takenaka and N.~Kawakami, J. Phys.: Conf. Ser. {\bfseries 400},  032099
  (2012).

\bibitem{Kubo2021}
K.~Kubo, Phys. Rev. B {\bfseries 103},  085118 (2021).

\bibitem{Slater1954}
J.~C. Slater and G.~F. Koster, Phys. Rev. {\bfseries 94},  1498 (1954).

\bibitem{Tang1998}
H.~Tang, M.~Plihal, and D.~L. Mills, J. Magn. Magn. Mater. {\bfseries 187},  23
  (1998).

\bibitem{Griffith1961}
J.~S. Griffith: {\em The {{Theory}} of {{Transition-Metal Ions}}} ({Cambridge
  University Press}, 1961).

\bibitem{Okabe1997}
T.~Okabe, J. Phys. Soc. Jpn. {\bfseries 66},  2129 (1997).

\bibitem{Bunemann1998}
J.~B{\"u}nemann, W.~Weber, and F.~Gebhard, Phys. Rev. B {\bfseries 57},  6896
  (1998).

\bibitem{Kobayashi2006}
K.~Kobayashi and H.~Yokoyama, Physica C {\bfseries 445--448},  162 (2006).

\bibitem{Kubo2009JPCS}
K.~Kubo, J. Phys.: Conf. Ser. {\bfseries 150},  042101 (2009).

\bibitem{Kubo2009PRB}
K.~Kubo, Phys. Rev. B {\bfseries 79},  020407(R) (2009).

\bibitem{Kubo2011JPSJSA}
K.~Kubo and P.~Thalmeier, J. Phys. Soc. Jpn. {\bfseries 80},  SA121 (2011).

\bibitem{Kubo2017JPSJ}
K.~Kubo and H.~Onishi, J. Phys. Soc. Jpn. {\bfseries 86},  013701 (2017).

\bibitem{Kubo2015JPCS}
K.~Kubo, J. Phys.: Conf. Ser. {\bfseries 592},  012039 (2015).

\bibitem{Kubo2015JPSJ}
K.~Kubo, J. Phys. Soc. Jpn. {\bfseries 84},  094702 (2015).

\bibitem{Kubo2015PP}
K.~Kubo, Physics Procedia {\bfseries 75},  214 (2015).

\bibitem{Momoi1998}
T.~Momoi and K.~Kubo, Phys. Rev. B {\bfseries 58},  R567 (1998).

\bibitem{Sakamoto2002}
H.~Sakamoto, T.~Momoi, and K.~Kubo, Phys. Rev. B {\bfseries 65},  224403
  (2002).

\bibitem{DeFranco2018}
C.~De~Franco, L.~F. Tocchio, and F.~Becca, Phys. Rev. B {\bfseries 98},  075117
  (2018).

\bibitem{Maurya2022}
A.~K. Maurya, M.~T.~H. Sarder, and A.~Medhi, J. Phys.: Condens. Matter
  {\bfseries 34},  055602 (2022).

\bibitem{Han1998}
J.~E. Han, M.~Jarrell, and D.~L. Cox, Phys. Rev. B {\bfseries 58},  R4199
  (1998).

\bibitem{Ono2003}
Y.~{\=O}no, M.~Potthoff, and R.~Bulla, Phys. Rev. B {\bfseries 67},  035119
  (2003).

\bibitem{deMedici2011PRB}
L.~{de' Medici}, Phys. Rev. B {\bfseries 83},  205112 (2011).

\bibitem{deMedici2011PRL}
L.~{de' Medici}, J.~Mravlje, and A.~Georges, Phys. Rev. Lett. {\bfseries 107},
  256401 (2011).

\bibitem{Facio2017}
J.~I. Facio, V.~Vildosola, D.~J. Garc{\'i}a, and P.~S. Cornaglia, Phys. Rev. B
  {\bfseries 95},  085119 (2017).

\end{thebibliography}

\end{document}